\documentclass[12pt,letterpaper,copyedit]{article}
\usepackage{graphicx,epsfig,amsmath,subfigure,overcite}
   
\begin{document}

\title{Doppler-free, Multi-wavelength Acousto-optic deflector 
        for two-photon addressing arrays of Rb atoms
         in a Quantum Information Processor }

\author{Sangtaek Kim,$^{1,*}$ Robert R. Mcleod,$^1$ Mark Saffman,$^2$ and Kelvin H. Wagner$^1$ \\
$^1$Optoelectronic Computing Systems Center, \\
Department of Electrical \& Computer Engineering,\\ 
University of Colorado at Boulder, Boulder, CO\\
$^2$Department of Physics, University of Wisconsin, Madison, WI\\
$^*$ Corresponding author: sangtaek@colorado.edu}

\maketitle

\begin{abstract}
We demonstrate a dual wavelength acousto-optic deflector (AOD) designed 
to deflect two wavelengths to the same angles by driving with two RF 
frequencies.  The AOD is designed as a beam scanner to address two-photon 
transitions in a two-dimensional array of trapped neutral $^{87}$Rb atoms 
in a quantum computer.  
Momentum space is used to design AODs that have the same diffraction angles 
for two wavelengths (780 nm and 480 nm) and have non-overlapping Bragg-matched 
frequency response at these wavelengths, so that there will be no crosstalk 
when proportional RF frequencies are applied to diffract the two wavelengths.  
The appropriate crystal orientation, crystal shape, transducer size, 
and transducer height are determined for an AOD made with a Tellurium 
dioxide crystal (TeO$_{2}$). 
The designed and fabricated AOD has more than 100 resolvable spots, widely 
separated bandshapes for the two wavelengths within an overall octave 
bandwidth, spatially overlapping diffraction angles for both 
wavelengths (780 nm and 480 nm), and a 4 $\mu$sec or less access time.  
Cascaded AODs in which the first device upshifts and the second downshifts 
allow Doppler-free scanning as required for addressing the narrow atomic 
resonance without detuning.  We experimentally show the 
diffraction-limited Doppler-free scanning performance and spatial resolution 
of the designed AOD.
\end{abstract}


\section{Introduction}\label{sec:intro}

\begin{figure}
\centerline{  
 \includegraphics[width=7.0cm, angle=-90]{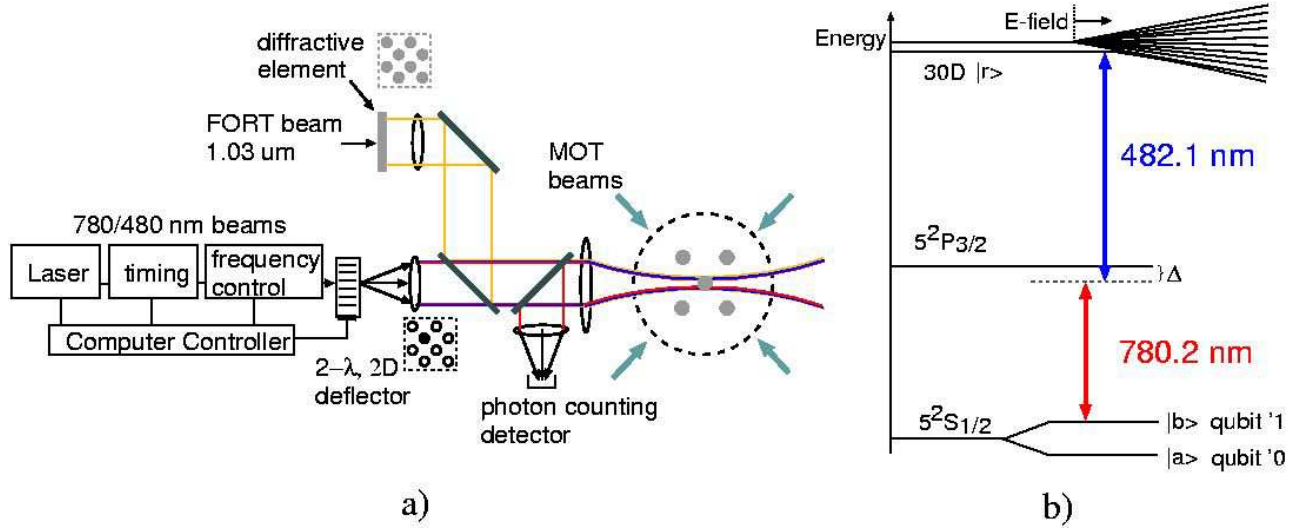}
}
\caption{a) Rydberg atom quantum processor 
with two-dimensional array of trapped atoms,  
b) Two-level Rydberg excitation using two wavelengths (780 nm/480 nm).  }
\label{fig:qtproc}
\end{figure}

Quantum algorithms may provide large gains in computational speed for 
certain problems such as factoring or database 
searching~\cite{Nielsen:00,Shor:97,Grover:97}.
Operation of quantum gates and execution of small quantum algorithms have 
been demonstrated using several different physical embodiments of quantum 
logic including nuclear magnetic resonance, cold trapped ions, single 
photons, and superconducting circuits~\cite{Vandersypen:ERSQFANMR,Leibfried:QDSTI,Plantenberg:DCQGSQB,Ralph:QOSIQIP}.  
The current challenge in experimental quantum computing is the task of 
scaling to larger numbers of qubits and logical operations. 
One of the implementations which appears particularly attractive because 
of its potential scalability is to use an array of neutral 
atoms trapped in optical lattices.  
Using well developed techniques of laser cooling and trapping, 
it is possible to create arrays of optical 
trapping sites and load them with single neutral atoms as shown in 
Figure~\ref{fig:qtproc} a).  
Long lived qubits can then be encoded in different hyperfine atomic ground 
states.  
Single qubit operations can be performed by stimulated transitions induced 
by laser beams focused on individual sites, or by Zeeman 
addressing~\cite{Schrader:NAQR,Jones:FQSCSTNA,Yavuz:06}.  
Several different physical mechanisms can potentially be used for 
two-qubit operations including collisions and dipole-dipole 
interactions~\cite{Brennen:QLGOL,Jaksch:FQGNA}.
High lying Rydberg levels with principal quantum number $n$ have dipole 
moments which scale as $n^2 e a_0$ where $e$ is the electronic charge 
and $a_0$ is the Bohr radius.  Dipole-dipole interactions of Rydberg atoms 
therefore scale as $n^4 e^2 a_0^2/r^3$, with $r$ the atomic separation.  
A detailed analysis of quantum gates using Rydberg interactions predicts 
the feasibility of high fidelity operations at MHz rates between optically 
resolvable sites by exciting Rydberg levels with 
$n\sim 70$~\cite{Saffman:04}. 

A peculiarity of the Rydberg scheme is that because the two-atom 
interaction utilizes excited states which have a finite radiative lifetime 
there is an optimum speed for which the gate fidelity is maximized.  
If the gate is run too slow radiative decay from the Rydberg states will 
lead to large errors, and if the gate is run too fast the dipole-dipole 
interaction will not be strong enough to provide the desired logical 
operation.  Analysis predicts\cite{Saffman:04} that gate 
speeds of several MHz, corresponding to sub-microsecond access times for 
individual qubits will be optimal.  Furthermore, transitions from the ground 
state to high lying Rydberg levels of alkali atom are in the deep UV 
spectral region where the unavailability and immaturity of lasers, 
modulators, detectors, and other optical technologies is inconvenient.  
A convenient technological alternative is to use a two-photon transition, 
which is resonantly enhanced by an intermediate level.  
For example in $^{87}$Rb ground to Rydberg transitions can be reached 
efficiently using a 780 nm laser that is 
near resonant with the $5s_{1/2} \rightarrow 5p_{3/2}$ transition followed 
by a 480 nm laser coupling to a Rydberg level such as 
$5p_{3/2}\rightarrow ns_{1/2}$ or $5p_{3/2}\rightarrow nd_{3/2,5/2}$ as 
shown in Figure~\ref{fig:qtproc} b). 
Two-photon Rabi oscillations between ground and Rydberg states using 
this approach have recently been demonstrated with single atoms in 
optical traps~\cite{Johnson:07}.

In light of the above considerations, implementation of a quantum processor 
based on a two-dimensional array of trapped atoms interacting via Rydberg 
states requires a fast beam scanner that provides $\mu\rm s$ access times 
for two different wavelengths. 
The required scanner specifications are that the output beam positions 
have to overlap for both wavelengths (780 nm and 480 nm), no Doppler shift 
of the sum of the two frequencies is tolerable (to avoid detuning of the 
two-photon transition), and the access time should be less than a few 
micro-seconds.  The number of qubits that can be addressed will be fixed 
by the number of resolvable spots of the scanner.  
Several different beamsteering technologies can be considered for this 
application.  Liquid crystal beam deflectors have millisecond scale access 
time, which is too slow~\cite{Wang:LCBGBD}.  Electro-optic prism beam 
deflectors are theoretically fast enough but require such high voltages 
that their speed is limited by the drive electronics to the milli-second 
range~\cite{Sun:PWPEBD}.  
MEMS mirrors are attractive since they work for both wavelengths without 
any Doppler shift, but sufficiently high-resolution beamsteering requires 
large apertures and deflection angles, which require milli-second response 
times, and are thus also too slow~\cite{Lin:FMOSTLOC}.  
A recent paper by Kim~\cite{Kim:DCMMITQC} demonstrated a MEMS mirror 
with 11 $\mu$s switching time between two sites, but this would be slower 
for addressing larger arrays. 
Acousto-optic deflectors (AODs) produce unwanted Doppler shifts but have  
advantages in terms of access time and resolution and can be crossed 
to scan two-dimensional arrays.  The Doppler shift can be canceled 
by cascading two AODs with opposite diffraction orders~\cite{Freyre:81} 
or by pre-compensation with a double-passed AOD that produces Doppler 
without any angular deflection.   
AODs are thus good candidates for two-dimensional, 
multi-color addressing because of their high access speed; however, the 
problems of Doppler shift and the difference of diffraction angles with 
wavelength are problems that must be solved to make use of AODs for this 
quantum information processing (QIP) application of two-color addressing of 
two-dimensional arrays of trapped atoms.  
This paper describes the design and development of 
a high-speed, high-efficiency, Doppler-free, multi-beam and multi-color 
acousto-optic scanner system usable in QIP applications to rapidly 
address arrays of trapped atoms. This type of device may also be of 
interest for other implementations of quantum logic such as trapped ions, 
superconductors, or quantum dots in semiconductors.

\section{Background on Acousto-optic devices}\label{sec:aod}

AODs consist of piezoelectric transducers bonded onto bulk photoelastic 
crystals such as TeO$_{2}$.  Applying RF electrical signals to the 
transducer launches bulk acoustic waves into the crystal, periodically 
modulating the crystal's dielectric tensor or index of refraction 
with a traveling wave.  
The modulation of the index of refraction functions as a volume index 
grating, which diffracts collimated light incident on the AOD at the Bragg 
angle.  The diffraction angle, $\theta \approx \frac{\lambda}{\Lambda}=
\frac{\lambda f}{V_{a}}$, is proportional to the applied RF 
frequency, $f$, and inversely proportional to the acoustic velocity, 
$V_{a}$, of the AOD.  The scan angle is also proportional to the optical 
wavelength $\lambda$, so the scan angles 
vary from $\theta_{1}\approx \frac{\lambda_{1} f}{V_{a}}$ to 
$\theta_{2}\approx \frac{ \lambda_{2} f}{V_{a}}$ 
for two different wavelengths given a fixed RF frequency.  
The wavelength-dependent scan angle of the AOD can be compensated 
by applying two RF frequencies simultaneously with the ratio of frequencies 
determined by the wavelengths 
($\frac{f_{1}}{f_{2}}=\frac{\lambda_{2}}{\lambda_{1}}$), so that 
each wavelength is diffracted into the same direction by the corresponding 
proportional RF frequency.
  
The number of resolvable spots of an AOD is given 
by time-bandwidth product ($N=T\cdot B$), where $T=A/V_{a}$ is the acoustic 
propagation time across the device aperture $A$, and $B$ is the bandwidth 
limited by the transducer acousto-electric conversion bandwidth and the 
Bragg-matched bandwidth, so one should be able to achieve better 
resolution with larger bandwidth AOD.  
However, since the acoustic attenuation increases with frequency 
as $\alpha=\alpha_{0} f^{2}$, the limiting device aperture decreases with  
frequency squared, thus the number of resolvable spots actually decreases 
with increasing frequency, so a better strategy is to lower the frequency 
to increase the $T\cdot B$ as long as a sufficiently large crystal is 
available.  
The acoustic velocity $V_{a}$ and the area of the beam incident on 
the AOD, $A$, determine the access time $T=\frac{A}{V_{a}}$ and 
resolution $N=\frac{A\cdot B}{V_{a}}$.  
Thus AO-addressing speed must be traded off with resolution.    

\begin{figure}
 \centerline{  
 \includegraphics[width=8.2cm, angle=-90]{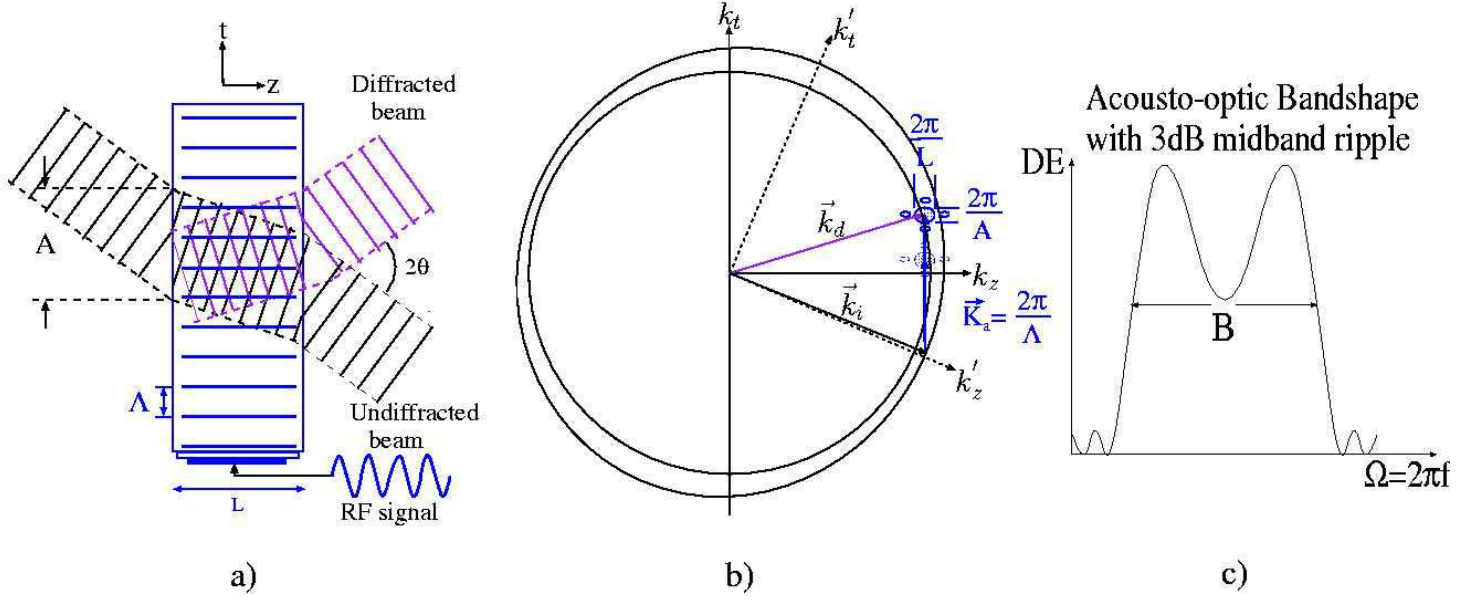}
}
\caption{ a) Real-space view of acousto-optic diffraction.  
A single frequency applied RF signal produces a sinusoidal grating inside 
the AO medium, which diffracts the incident plane wave with an 
angle 2$\theta$.  
b) Momentum space representation of acousto-optic diffraction for an 
acoustically and optically rotated TeO$_{2}$ device.  
As the frequency increases, the coupling 
efficiency between the uncertainty distribution and optical momentum 
surface gives the RF bandshape of the AOD.  
c) The acousto-optic bandshape with slight angle detuning gives the 
3 dB ripple bandshape.  }
\label{fig:mosurf}
\end{figure}  
%

As shown in Figure~\ref{fig:mosurf}, we can describe the diffraction of 
an AOD in momentum space (Fourier space), which provides a graphical 
picture of the acousto-optic interaction for our AOD design in the small 
diffraction efficiency limit (Born approximation). 
The Born approximation simplifies complex coupled-mode problem to linear 
equations allowing the use of graphical momentum space design 
techniques for optimizing the acousto-optic interaction geometry, 
even though in application we will use the devices with much higher 
diffraction efficiency~\cite{Weverka96:LLAOPS}.

The simplest description of the requirement for the conservation of 
momentum in acousto-optic diffraction requires the closure of the triangle 
of the input optical wavevector, $\vec{k}_{i}$ (which lies on the optical 
momentum surface), with the acoustic grating wavevector, $\vec{K}_{a}$, to 
give the output optical wavevector, $\vec{k}_{d}$.  
Often, slight deviations 
from perfect momentum matching are attributed to momentum mismatch 
$\Delta \vec{K}$ describing the deviation of $\vec{k}_{i} + \vec{K}_{a}$ 
from $\vec{k}_{d}$~\cite{Yariv:03}.  Instead, to account for slight 
allowable momentum mismatch in finite media, we introduce the concept of 
momentum uncertainty, given by the 3-D Fourier transform of the 
interaction region and enforce the requirement that the diffracted field is a 
superposition of plane wave components lying on the optical momentum 
surface~\cite{Weverka96:LLAOPS,Mcleod95:SADTOS}.  
The convolution of the 3-D Fourier transform of the incident optical field 
with the dielectric perturbation caused by the acoustic wave launched by 
the transducer (acoustic grating) sampled by the $\vec{k}$-sphere of the 
allowed propagating eigenmodes represents the angular spectrum of the 
diffracted optical wave.  
When the transducer is driven by frequency $\Omega$, it launches a bulk 
acoustic wave causing a time-varying impermeability tensor 
\begin{eqnarray} 
\Delta{\underline{\underline{\eta}}}(\vec{r},t) & = & \frac{1}{(2\pi)^{2}} \int \int \underline{\underline{\underline{\underline{p}}}} \underline{\underline{\hat{S}}}(K_{y}, K_{z}) (\underline{\underline{\hat{S}}}_{T} \cdot \underline{\underline{\hat{S}}}(K_{y}, K_{z})) A(K_{y}, K_{z})  \\
&   &\times\frac{1}{2\pi} \int  R(\Omega) {\rm exp}[ i( \Omega t - (\sqrt{ [\Omega/V_{a}(K_{y},K_{z})]^{2}-K_{y}^{2}-K_{z}^{2} }x + K_{y}y + K_{z}z )) ] d\Omega dK_{y}dK_{z},  \nonumber
\label{eq:diepert} 
\end{eqnarray}
where $\underline{\underline{\underline{\underline{p}}}}$ is the photoelastic 
tensor, $\underline{\underline{\hat{S}}}(K_{y}, K_{z})$ is the 
directional-dependent acoustic strain polarization tensor while 
$\underline{\underline{\hat{S}}}_{T}$ is the strain polarization produced 
by the transducer, 
$R(\Omega)$ is the Fourier transform of the applied RF signal, 
$A(K_{y},K_{z})$ is the Fourier transform of the transducer aperture function, 
$a(y,z)$. 
In the case of TeO$_{2}$ crystal, 
$\underline{\underline{\hat{S}}}(K_{y}, K_{z})$ is approximately equal to 
[1\={1}0] for all directions near [110], so that $\underline{\underline{\hat{S}}}_{T} \cdot \underline{\underline{\hat{S}}}(K_{y}, K_{z})$ is approximately 
equal to 1.  
The dielectric tensor perturbation is given by 
\begin{equation}
\Delta{\underline{\underline{\epsilon}}}(\vec{r},t) = 
\underline{\underline{\epsilon}} \Delta{\underline{\underline{ \eta}}} (\vec{r},t) \underline{\underline{\epsilon}}, 
\label{eq:diper}
\end{equation}
where $\underline{\underline{\epsilon}}$ is the permittivity tensor.    
We can represent the angular 
spectrum of the diffracted $E_{d}(\vec{r},t)$ optical wave as:
\begin{equation}
  E_{d}^{\omega_{d}}(k_{x},k_{y},L)\,
 =\frac{i\omega_{d}^{2}}{2c^{2}k^{q}_{zd}(k_{x},k_{y})}\,
  \int \delta(k_{z}-k^{q}_{zd}(k_{x},k_{y}))\,
 \mathcal{F}_{\vec{r}}\bigg\{ \Delta {\underline{\underline{\epsilon}}} (\vec{r},\Omega) \vec{E}^{*}_{i}(\vec{r},\omega)\bigg\}\cdot \hat{p}_{q}(k_{x}, k_{y})dk_{z}\,
\label{eq:aomosp9} 
\end{equation}
where $c$ is the velocity of light, $L$ is the crystal interaction length 
(transducer length), 
$k^{q}_{zd}(\vec{k}_{t})=\sqrt{k_{d}^{2}-|\vec{k}_{t}|^{2}}$ 
is the longitudinal component of the diffracted wavevector solved from an 
eigenvalue problem for the homogeneous crystal with corresponding 
polarization eigenvectors $\hat{p}_{q}(k_{x}, k_{y})$ for the two allowed 
modes.
$k_{d}=2\pi n_{d}(\vec{k}_{t},\omega_{d})\omega_{d}/c$ describes the 
anisotropic magnitude of the diffracted wavevector, 
$\vec{k}_{t}=\hat{x}k_{x}+\hat{y}k_{y}$ is the 
transverse component of the wavevector, and $\vec{E}_{i}(\vec{r},\omega)$ 
is the incident optical field at the crystal boundary with optical 
frequency $\omega$.  The Doppler shift of the diffracted beam is given by 
the conservation of energy $\omega_{d}=\omega \pm \Omega$ corresponding to 
Doppler upshifting (+) and downshifting (-) orders.  
In this expression, the 3-D Fourier transform of the product of the incident 
field and the dielectric perturbation is sampled by the delta function 
representing the momentum surface of allowed propagating modes, which give 
the source term for the angular spectrum of the diffracted field at the 
output of AO medium.  

We use momentum space (or $\vec{k}$-space) as summarized by 
Equation~\ref{eq:aomosp9} to design the dual-wavelength 
AODs required for this QIP application.  
When the piezoelectric transducer launches an acoustic wave, a grating 
is produced through the photoelastic effect.
The grating is represented in $\vec{k}$-space as a grating vector whose 
length $|\vec{K}_{a}|=\Omega/V_{a}$ with an uncertainty distribution at 
its tip due to the finite extent of the acoustic wave, where $\Omega$ is 
the angular radian RF frequency.  
We can regard the propagating acoustic wave as a moving grating with 
frequency $\Omega$ (grating period $\Lambda=2\pi V_{a}/\Omega$) and 
grating width $L$ (given by the transducer length).   
The incident electric field ($E_{i}(\vec{r},\omega)$) has Bragg-matched 
wavevector $|\vec{k}_{i}|=\frac{2\pi n_{i}}{\lambda}$ and we approximate 
its envelope as a rectangle of width $A$ in the t direction, as shown 
in Figure~\ref{fig:mosurf} a).  
As shown by the term in braces in Equation~\ref{eq:aomosp9}, 
the real-space product of the dielectric perturbation 
$\Delta \epsilon(\vec{r},\Omega)$ and the incident electric field 
yields the material polarization that has a rectangular profile 
of $A \times L$ as shown in Figure~\ref{fig:mosurf} a).  
Its Fourier transform is therefore a product of sinc function of width 
$2\pi/A \times 2\pi/L$ displaced from the origin by a carrier 
$\vec{k}_{i} + \vec{K}_{a}$, as shown in Figure~\ref{fig:mosurf} b).  
This Fourier uncertainty distribution represents the 
acousto-optically induced polarization responsible for diffracting the 
optical field as represented in momentum space.  
The uncertainty distribution is illustrated as a single contour line, 
which is the product of sinc functions rendered in momentum space shown in 
Figure~\ref{fig:mosurf} b).  
In general, this uncertainty distribution is represented with a 3-D 
Fourier transform of the product of the incident optical wave and 
acoustic field amplitude.  
The angular distribution of the diffracted field is then found by 
sampling this uncertainty distribution with the allowed propagating 
eigenmode as shown in Equation~\ref{eq:aomosp9} and 
Figure~\ref{fig:mosurf} b).
When the RF frequency $\Omega=2 \pi f$ is applied, the acoustic 
momentum $\vec{K}_{a}$ linearly increases and tangentially skims 
along the diffracted optical momentum surface with slight angular 
detuning.  
The acousto-optic bandshape versus applied RF frequency can be found 
geometrically in momentum space as the overlap of the uncertainty 
distribution with the allowed propagating eigenmodes as 
illustrated in Figure~\ref{fig:mosurf} b).
The resulting 3 dB rippled acousto-optic bandshape is as illustrated in 
Figure~\ref{fig:mosurf} c)~\cite{Mcleod:93,Weverka96:LLAOPS}.

\begin{figure}

\centerline{  
 \includegraphics[width=8.0cm, angle=-90]{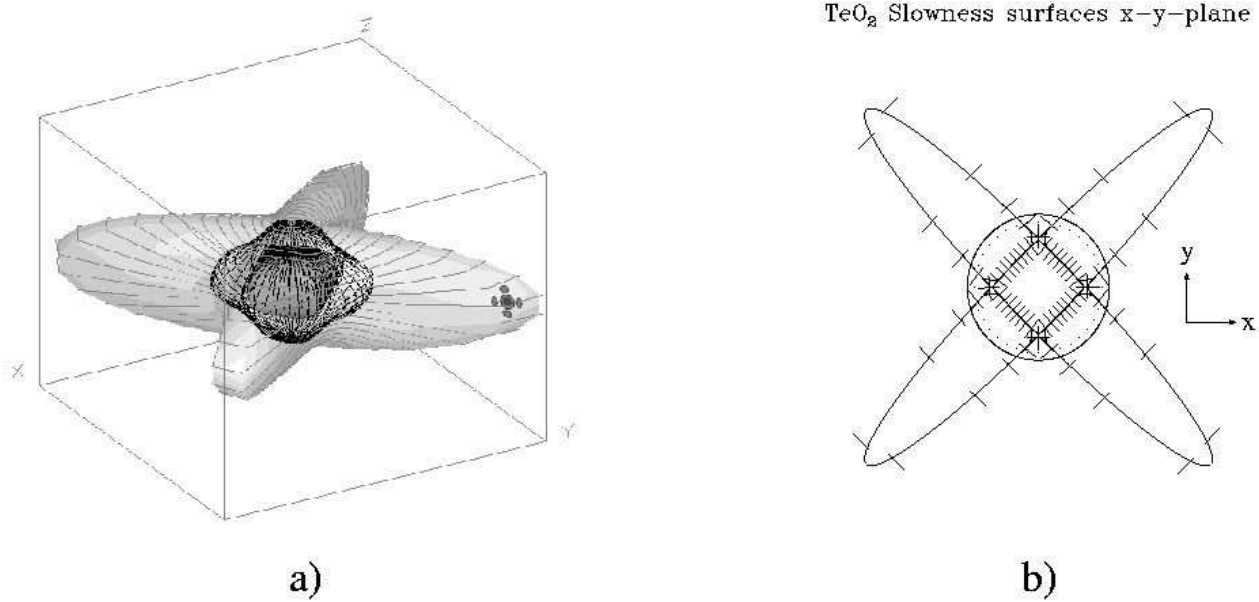} }
\caption{ a) Three-dimensional TeO$_{2}$ slowness surfaces with 
3 different acoustic polarizations and the transducer Fourier transform 
$P(K_{y},K_{z})$ projected onto the surface for an acoustically rotated 
geometry~\cite{Mcleod95:SADTOS}.  
b) Top view of TeO$_{2}$ 
slowness surface.  Along [110], the acoustic velocity is extremely slow 
which results in high diffraction efficiency.  
The indicated slow shear wave acoustic polarization remains nearly constant 
as the propagation direction deviates from [110]. }
\label{fig:teo2}
\end{figure}    
%

For our AOD beam scanner design, we chose TeO$_{2}$ as our AO crystal 
because of its anomalous slow shear mode with high diffraction efficiency 
and high resolution when compared to other 
crystals (e.g., GaP or LiNbO$_{3}$)~\cite{Uchida:71}.  
Using the stiffness tensor coefficients of the TeO$_{2}$ crystal, 
the directional dependent acoustic velocity can be determined from the 
eigensolution of the Christoffel equation~\cite{Auld:86,Jieping:92}.  
The inverse velocity, or slowness surface, is shown in 
Figures~\ref{fig:teo2} a) and b).
Along the [110] directions, the velocity of the shear wave 
polarized in the \={X}Y plane (0.62 mm/$\mu$sec) is nearly 10 times slower 
than in conventional AO crystals.  The slow acoustic velocity of 
TeO$_{2}$ gives a high figure of 
merit $M_{2}=(n^{6}p^{2})/(\rho V_{a}^{3})$ ($n$ is refractive index, 
$p$ is the photo-elastic constant, $\rho$ is the material density) which 
is the key material parameters that determines acousto-optic diffraction 
efficiency $\eta$. 
\begin{equation}
 \eta \equiv \frac{I_{d}}{I_{0}}=\sin^{2} \left[ \frac{\pi^{2}P_{a}L}{2\lambda^{2}H}M_{2} \right]^{1/2}, 
\label{eq:deff} 
\end{equation}
where $P_{a}$ is the acoustic power, $H$ is the transducer height, and $L$ is 
the transducer length.  
The slow velocity of TeO$_{2}$ not only yields high $M_{2}$ and high 
efficiency, but also enables high resolution and large angular diffraction 
at moderate drive frequencies that yield large time-bandwidth product 
operation.  However, the large acoustic attenuation of 
18 dB/$\mu$s$\cdot$GHz$^{2}$, the slow access speed, and the large walk-off 
angles for acoustically rotated devices present challenges to the use of 
this slow-shear mode in TeO$_{2}$~\cite{Uchida:71,Auld:86,Jieping:92}. 

The acoustic walk-off angle is determined by the tilt of the acoustic 
slowness surface relative to the acoustic grating vector ($\vec{K}_{a}$).  
Second and higher-order terms in the Taylor expansion of the slowness 
surface represent acoustic diffraction.  
This diffraction is quantified by the Rayleigh-range that indicates the 
well-collimated region (distance $Z_{0}$) of the diffracting wave, 
which is defined as 
\begin{equation}
 Z_{0}=\frac{D^{2}}{b\Lambda_{a}}, 
\label{eq:raylegh} 
\end{equation}
where $D$ is the spatial width of propagating wave, $\Lambda_{a}$ is the 
wavelength, and $b$ is inverse of the slowness surface radius of curvature 
relative to an isotropic material~\cite{Mcleod95:SADTOS}. 
For an acoustic wave propagating along the [110] direction in TeO$_{2}$ 
crystal, the diffracting power (or excess curvature) in $z-xy$ plane 
($b_{z}$) is 11, and 52 for the $xy$($b_{t}$) plane.
The walk-off angle can be approximately given by $b\cdot\theta$ 
for small $\theta$, where $\theta$ is the angle between the acoustic wave 
propagation direction and the symmetry axis direction 
([110] in TeO$_{2}$)~\cite{Jieping:92}.  

\begin{figure}
\centerline{  
 \includegraphics[width=8.2cm, angle=-90]{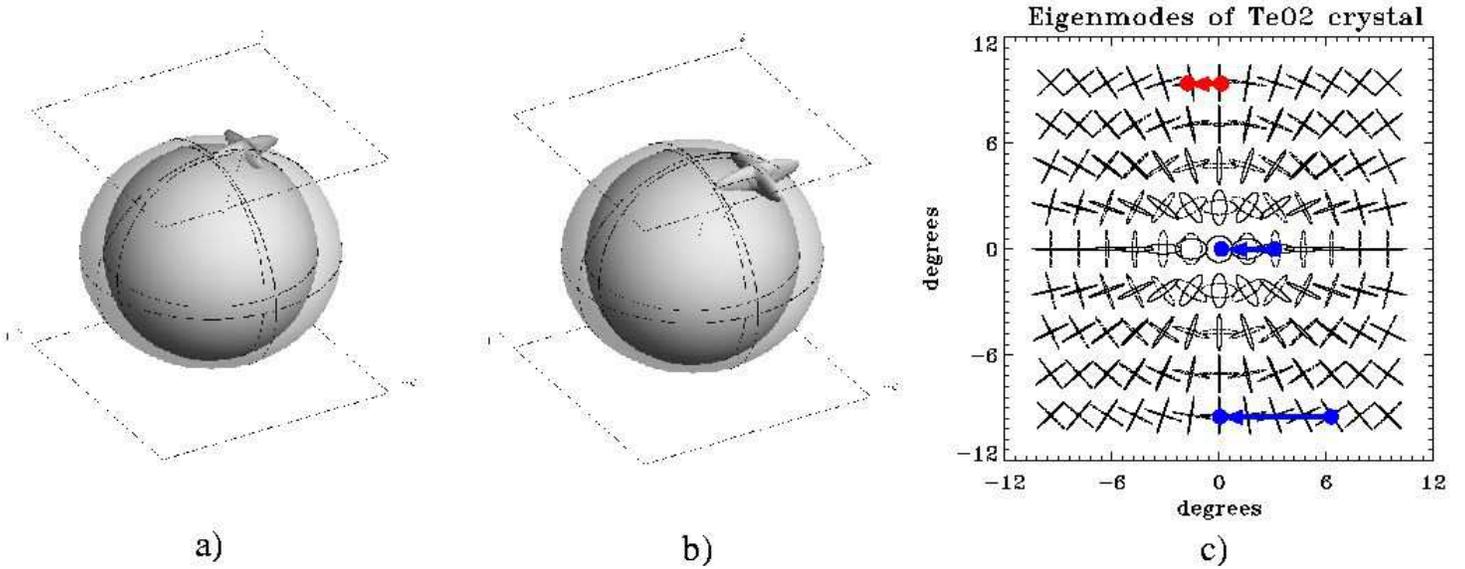}
}
\caption{ a) 3-D momentum space showing the tangentially 
phase-matched acousto-optic interaction around the z-axis with an exaggerated 
splitting due to optical activity and a corresponding exaggerated acoustic 
frequency~\cite{Mcleod95:SADTOS}.  
b) Optically rotated acousto-optic interaction in TeO$_{2}$.  
c) The eigen polarization of a TeO$_{2}$ crystal near the optic axis 
for a wavelength of 780 nm over a range of $\pm$ 10 degrees.  The eigen 
polarization is circular near the optic axis and changes to linear 
polarization as the angle deviates by a few degrees away from 
the optic axis. }
\label{fig:optteo2}
\end{figure}    
%

TeO$_{2}$ is a positive uniaxial crystal with a slight splitting between 
the eigen surfaces along the z-axis due to optical activity.  
This splitting of the eigen surfaces allows tangential birefringent 
diffraction near the z-axis, since the acoustic slow-shear mode launched 
from the transducer can couple light from the outer extraordinary 
eigen surface to the inner ordinary eigen surface~\cite{Uchida:71}.  
The momentum-matched interactions are illustrated by the intersections 
between the acoustic momentum surface centered at the $\vec{k}_{e}$ vector 
of the incident light with the inner optical momentum surface, as 
illustrated in three-dimension in Figure~\ref{fig:optteo2} a), with an 
exaggerated optically active splitting and correspondingly increased 
acoustic frequency for illustrative clarity.  
Optical rotation is a rotation around the normal to the transducer 
that rotates to a plane with larger splitting between the eigen surfaces 
and thus requires a higher momentum matching frequency and correspondingly 
larger acoustic momentum as illustrated in Figure~\ref{fig:optteo2} b).  
Optical modes propagating near the z-axis have 
circular polarization eigenmodes, while propagation directions further from 
the z-axis, as illustrated in Figure~\ref{fig:optteo2} c), have eigenmodes 
that approach linear polarization for rotation of a few degrees away from 
the z-axis. 
Without optical rotation, an elliptically polarized incident wave diffracts 
to a circularly polarized eigenmode as shown in the middle part of 
Figure~\ref{fig:optteo2} c).  With a few degrees of optical rotation, 
the incident and diffracted eigenmodes become nearly linear and a larger 
acoustic momentum is required for tangential on-axis diffraction as shown 
at bottom of Figure~\ref{fig:optteo2} c).  
The required acoustic momentum and its eigenmodes for the acoustically 
rotated AODs designed in section~\ref{sec:design} are also illustrated in 
the upper part of Figure~\ref{fig:optteo2} c).

\begin{figure}
 \centerline{  
 \includegraphics[width=5.0cm, angle=-90]{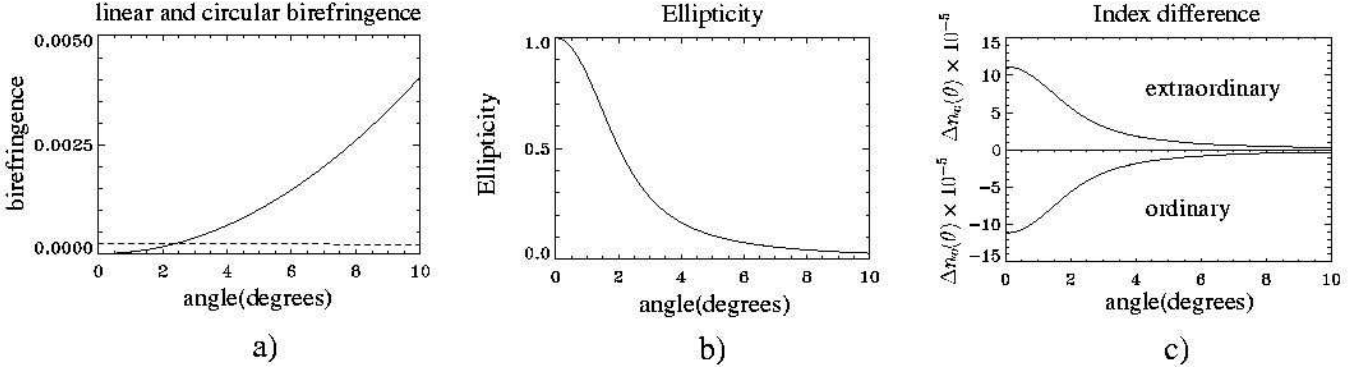}
}
\caption{ a) The solid line is the linear birefringence 
$\Delta n_{l}$ and the dashed line is the circular birefringence 
$\Delta n_{c}$ at 780 nm wavelength.  
b) The ellipticity $\xi$ as a function of angle from the z-axis.
c) The difference of the TeO$_{2}$ indices of refraction when there is 
optical activity and when there is no optical activity as a 
function of an angle $\theta$ from the z-axis.  }
\label{fig:noptact}
\end{figure}  

The $\hat{o}$ and $\hat{e}$ eigenmodes have circumferential and radial 
major axis orientations that are most easily expressed using spherical 
polar coordinates.  
Ellipticity ($\xi$) is defined as the ratio of the major ($a$) and 
minor ($b$) axes ($\xi=\frac{b}{a}$) of the polarization ellipse, 
giving $\pm$1 for circular polarization and 0 for linear polarization.  
For a propagating angle $\theta$ away from the z-axis, 
the ellipticity of the incident and diffracted eigenmodes is given 
by~\cite{Jieping:92} 
\begin{equation}
  \xi=\frac{\Delta n_{c}(\theta)}{\Delta n(\theta)+\Delta n_{l}(\theta)},
\label{eq:ellipt} 
\end{equation}
where $\Delta n(\theta)=\sqrt{[\Delta n_{c}(\theta)]^{2}+
[\Delta n_{l}(\theta)]^{2}}$ is the total birefringence at an angle 
$\theta$ away from the z-axis,  
$\Delta n_{l}(\theta)=n_{e}(\theta)-n_{o} \approx (n_{e}-n_{o})\sin^{2}\theta$ 
is the linear birefringence when optical activity is neglected, 
and $\Delta n_{c}(\theta) \approx \frac{\rho_{r} \lambda}{\pi}\cos^{2}\theta$ 
is the circular birefringence induced by optical activity near the optical 
axis, where $\rho_{r}$ is the 
optical rotatory power defined as the rotation angle of the linear 
polarization per unit length. 
The circular birefringence ($\Delta n_{c}$) is dominant close to the z-axis, 
giving circular optical eigenmodes along the z-axis, 
while the linear birefringence ($\Delta n_{l}$) dominates further away 
from the z-axis, eventually giving nearly linear 
optical eigenmodes polarization for more than 10 degrees away from the 
z-axis as shown in Figure~\ref{fig:noptact} a).  At 780 nm wavelength, 
the ellipticity is $\pm$1 along the z-axis, and decreases to 0.027 when 
the propagation direction is 10 degrees away from the z-axis as shown in the 
Figure~\ref{fig:noptact} b).  The optical activity changes the ordinary and 
extraordinary index of refraction around the z-axis.  
The difference between an ordinary index of refraction with optical 
activity and one without optical activity ($n_{o}(\theta)-n_{o}$) and 
the extraordinary difference with and without optical activity for 
780 nm wavelength are illustrated in Figure~\ref{fig:noptact} c).  
Along the z-axis, the deformed ordinary index surface 
(or momentum surface, $k=2\pi n /\lambda$) is 
pushed in, and the extraordinary surface (or momentum surface) is dimpled 
out as shown in Figure~\ref{fig:noptact} c)~\cite{Fukumoto:75}.

\section{Feasibility demonstration using off-the-shelf devices}\label{sec:demo}

\begin{figure}
 \centerline{  
 \includegraphics[width=14.0cm, angle=-90]{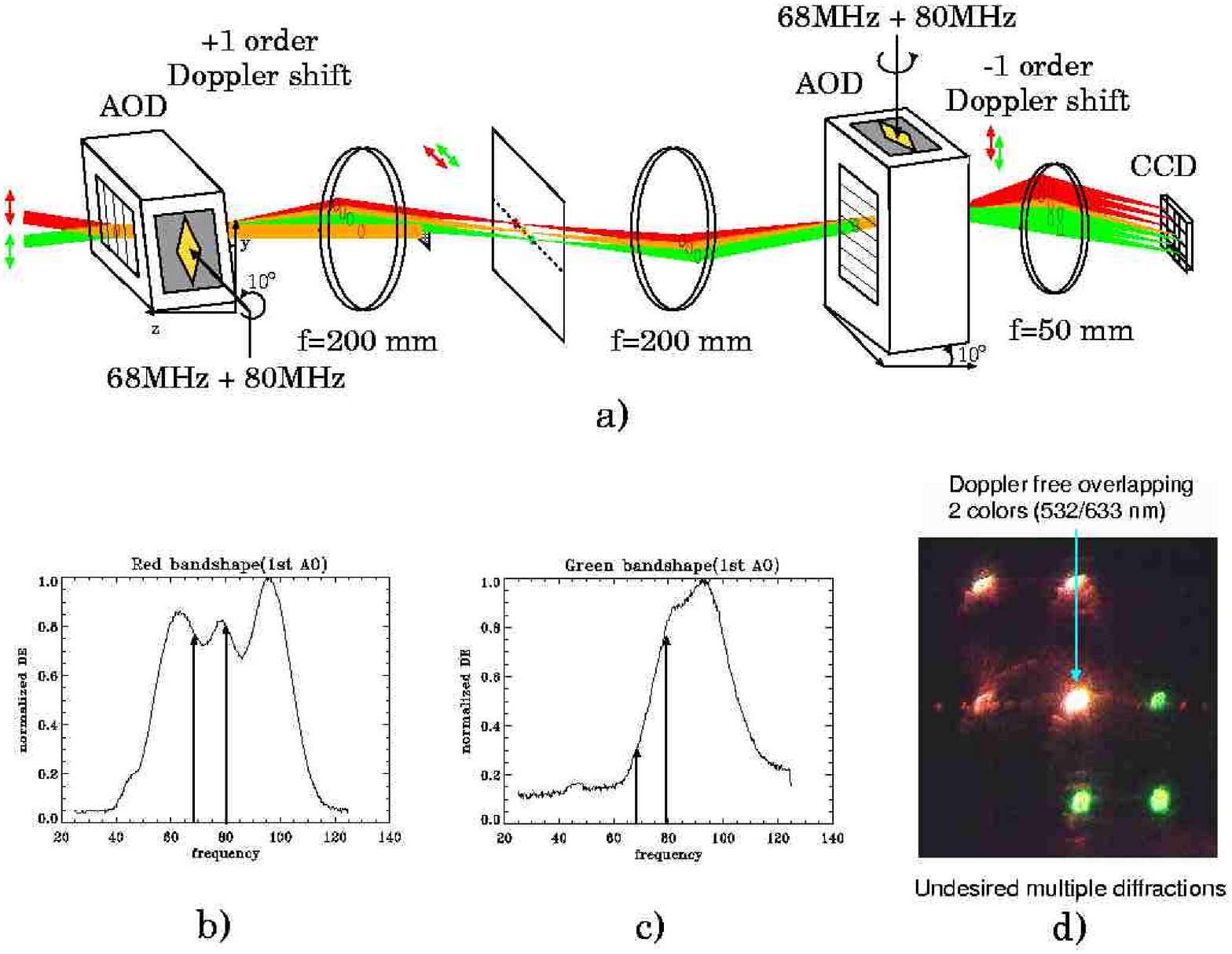}
}
\caption{ a) The experimental setup for 
demonstrating multi-color 
Doppler-free AO diffraction at two wavelengths (633 nm/532 nm).  b) The RF 
bandshape of the TeO$_{2}$ AOD at 633 nm using tangential phase matching 
geometry.  c) The bandshape of the TeO$_{2}$ AOD at 532 nm.  The center 
frequency is higher at the shorter wavelength.  d) The diffracted spots 
from the second AOD when driven by the two RF frequencies.  The spot at the 
center is the deflected Doppler-free position where the two wavelengths 
overlap, while the other six are undesired cross diffractions.}
\label{fig:demo}
\end{figure}

As an initial demonstration, we performed multi-color Doppler-free 
beamsteering with a 633 nm HeNe laser and a 532 nm Ar laser using two 
off-the-shelf, conventional AODs, as shown in Figure~\ref{fig:demo}.  
The AODs are made of TeO$_{2}$ crystals and are oriented by rotating 
about the acoustic $\vec{K}_{a}$ vector (transducer normal) with a 
10-degree optical-rotation 
angle away from the z-axis (which is found by observing the conoscopic 
pattern using converging light between crossed polarizers).  
The eigenmode polarization of the input optical wave is rotated about 
6 degrees from the vertical axis and the eigenmode of the output optical 
wave is horizontally oriented.  For this large optical rotation angle, 
the ellipticity of the eigenmodes is about 3\%, so appropriately 
oriented linear polarizations will achieve about 97\% coupling efficiency 
to the eigenmode.  
Figures~\ref{fig:demo} b) and c) show the bandshapes of the AOD for two 
different wavelengths (633 and 532nm), both of which satisfy the tangential 
phase-matching condition.  The 3 dB bandwidth is 50 MHz (50-100MHz) for 
633 nm light and 30 MHz (75-105MHz) for 532 nm light.  
There is an overlap of the bandshapes for both wavelengths that ranges 
from 75 to 100 MHz that will result in extra unwanted diffraction of the 
wrong wavelength.  

We need to apply two frequencies (combined with an RF adder) to address 
a single spot with both wavelengths.  We can compensate for the angular 
difference caused by the wavelength difference using two RF frequencies 
with a ratio given by the inverse wavelength ratio, 
$\frac{f_{1}}{f_{2}}=\frac{\lambda_{2}}{\lambda_{1}}$. 
The first AOD, which is driven by two frequencies (68 MHz and 80 MHz) and 
oriented with a Doppler-upshifting diffraction, was imaged onto a second, 
90-degree rotated, Doppler-downshifting AOD driven by the same two 
frequencies, canceling the Doppler shift and giving a linear scan 
along a 45$^{\circ}$ tilted line.  
The 90-degree rotation places the angular scan from the first AOD 
in the plane of Bragg degeneracy of the second AOD to avoid 
bandwidth-restricting Bragg-mismatching effects~\cite{Guilfoyle:PITD}.  
With appropriate RF frequencies, Doppler-free overlapping diffracted 
spots are produced, as shown in Figure~\ref{fig:demo} d).  
However, undesired diffractions are also produced because the bandshape 
for the two wavelengths overlap in the RF frequency domain.  
One of our design goals is to get rid of these undesired 
diffractions thereby improving the achievable efficiency of the desired 
Doppler-compensated diffractions and producing no unwanted 
extraneous terms.

\section{Custom AOD Design}\label{sec:design}

\begin{figure}
 \centerline{  
 \includegraphics[width=6.5cm, angle=-90]{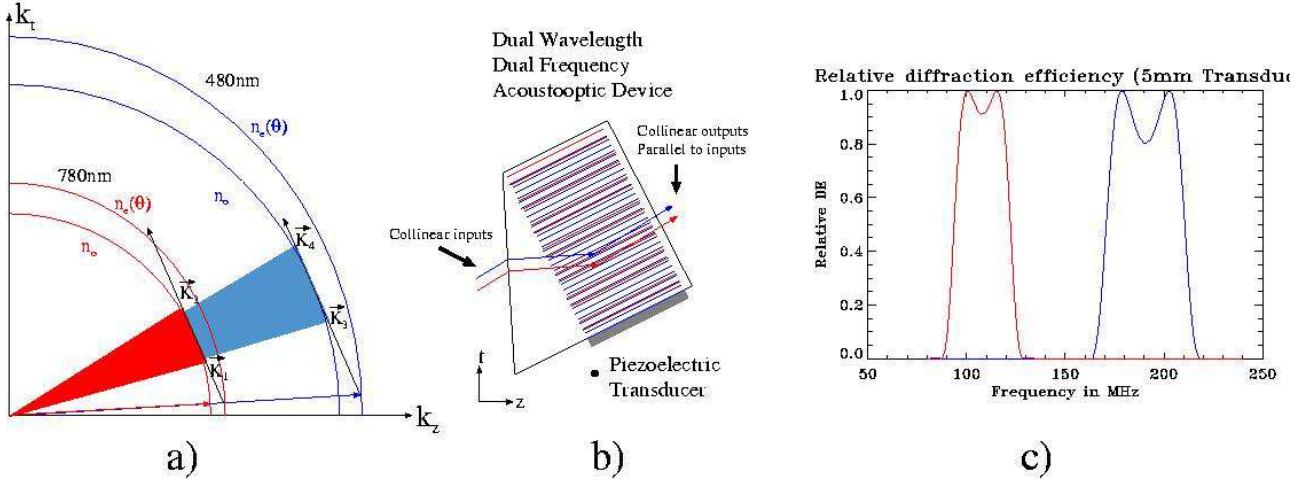}
}
\caption{ a) The momentum-space conceptual 
design of two-color spatially overlapping diffraction using anisotropic 
diffraction.  b) Real-space design for the dual RF 
frequency input acousto-optic deflector for two-color diffraction.  
The AOD prism cut allows Bragg matching for parallel two-color inputs 
and yields an undeviated midband output for both colors.  
c) Dual bandshape for the two 
wavelengths are completely separated, enabling only a single overlapping 
Doppler-free deflection to be produced at both colors simultaneously. }
\label{fig:kdesign}
\end{figure}

For this spatially-overlapping, two-color diffraction with the largest 
efficiency-bandwidth product, we employ tangentially phase-matched 
anisotropic diffraction for both wavelengths simultaneously, 
as shown schematically in Figure~\ref{fig:kdesign} a).
The momentum surface for 480 nm is scaled up from that at 780 nm by the 
ratio of material wavelengths, and the 
actual splittings between the ordinary and extraordinary surfaces are 
exaggerated, but the basic geometry of the interaction is illustrated.  
The tangential geometry with maximal efficiency-bandwidth product is 
achieved when the correct polarization is incident at an angle onto the 
outer momentum surface for which the sum of $\vec{k}_{i}+\vec{K}_{a}(f)$ 
skims tangentially across the inner momentum surface as the frequency is 
varied, so the phase mismatch is minimized across a wide bandwidth.  
The incident light at each color is tangentially diffracted by a range 
of non-overlapping acoustic frequencies towards a common overlapping 
region of output angles, and the transducer is lengthened compared to a 
conventional transducer to give suboctave bandwidth for each color.  
The full range of frequencies is contained 
in a single octave to avoid second harmonic diffractions and simplify the 
transducer impedance matching.  
For easy alignment of the system, a prism cut is 
utilized at the front face of the TeO$_{2}$ crystal, 
which allows parallel collimated beams 
of both wavelengths to refract into the required Bragg-matching input 
angles, and at the output face to leave the diffracted beams undeviated 
at the midband for each color. 

TeO$_{2}$ is an optically active, positive uniaxial crystal, which allows 
polarization-switching tangential birefringent diffraction at a 
convenient, low RF frequency. 
For conventional TeO$_{2}$ deflectors, the Bragg-matched interaction takes 
place near the z-axis of the crystal, where the slight splitting of the 
eigen surfaces due to optical activity is observable.  
For this dual-wavelength quantum computing application, the tangentially 
phase-matching frequency range for both wavelengths (780 nm/480 nm) was 
examined for both optical and acoustic rotation to find an optimal geometry.
Various optically rotated planes of incidence were examined, and 
in these optically rotated planes of incidence, various acoustic rotations 
were analyzed to find the optimum crystal orientation.  
Optical rotations about the transducer face normal correspond to tilts 
out of the interaction plane while acoustic rotations correspond 
to tilts of the transducer in the plane of the AO interaction.
The spacing between the inner and outer optical momentum surface increases 
with either optical or acoustic rotations, which requires larger RF 
frequencies for tangential phase matching.  

%
\begin{figure}
 \centerline{  
 \includegraphics[width=7.7cm, angle=-90]{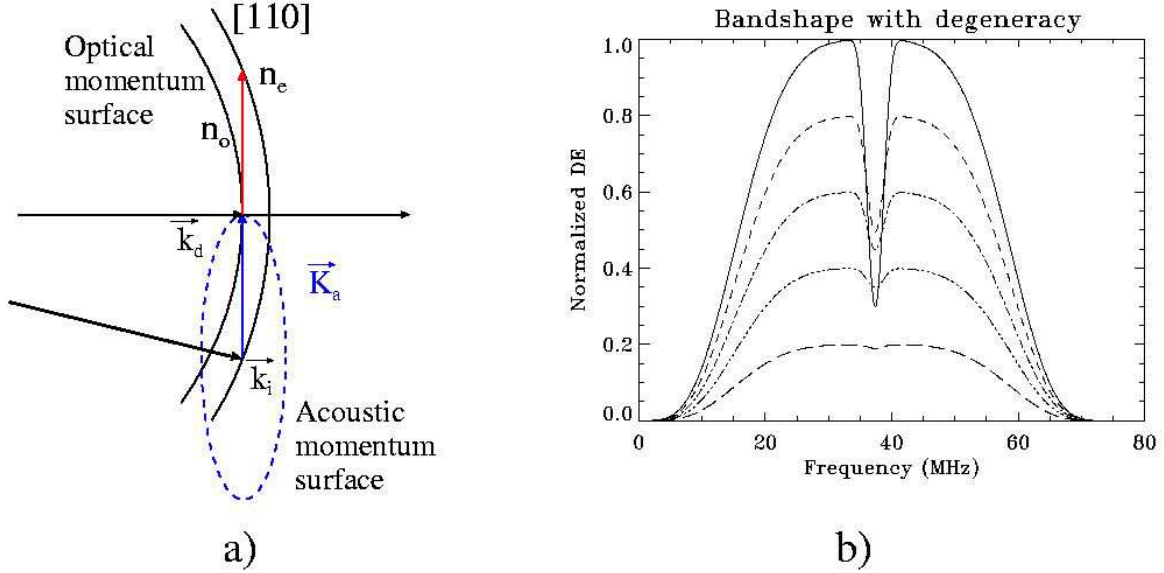}
}
\caption{ a) Optical momentum space showing midband degenerate 
diffraction.  The size of acoustic momentum for the extraordinary outer 
momentum surface to the ordinary inner momentum surface for tangential 
birefringent diffraction is equal to the ordinary inner momentum surface 
to the extraordinary outer momentum surface diffraction due to the 
symmetry of optical momentum surface.  
b) As applied RF power increases, the ordinary inner momentum surface 
to the extraordinary outer momentum surface diffraction causes a dip 
in the middle of AO bandshape. }
\label{fig:midegen}
\end{figure}  

Figure~\ref{fig:midegen} a) shows a slice of the symmetric optical momentum 
surface and acoustic momentum $\vec{K}_{a}$ satisfying tangential birefringent 
diffraction from the extraordinary outer momentum surface to the ordinary 
inner momentum surface.   Due to the symmetry of this geometry, the acoustic 
momentum for tangential birefringent diffraction 
($\vec{K}_{a}$) near the central frequency satisfies the second order 
diffraction condition from the ordinary inner momentum surface to the 
extraordinary outer momentum surface, 
referred to as midband degeneracy~\cite{Warner:72}.  
As the applied RF power increases for higher diffraction efficiency, 
the midband degeneracy causes a large dip in the 
middle of the AO bandshape, as shown Figure~\ref{fig:midegen} b).

%
\begin{figure}
 \centerline{  
 \includegraphics[width=8.9cm, angle=-90]{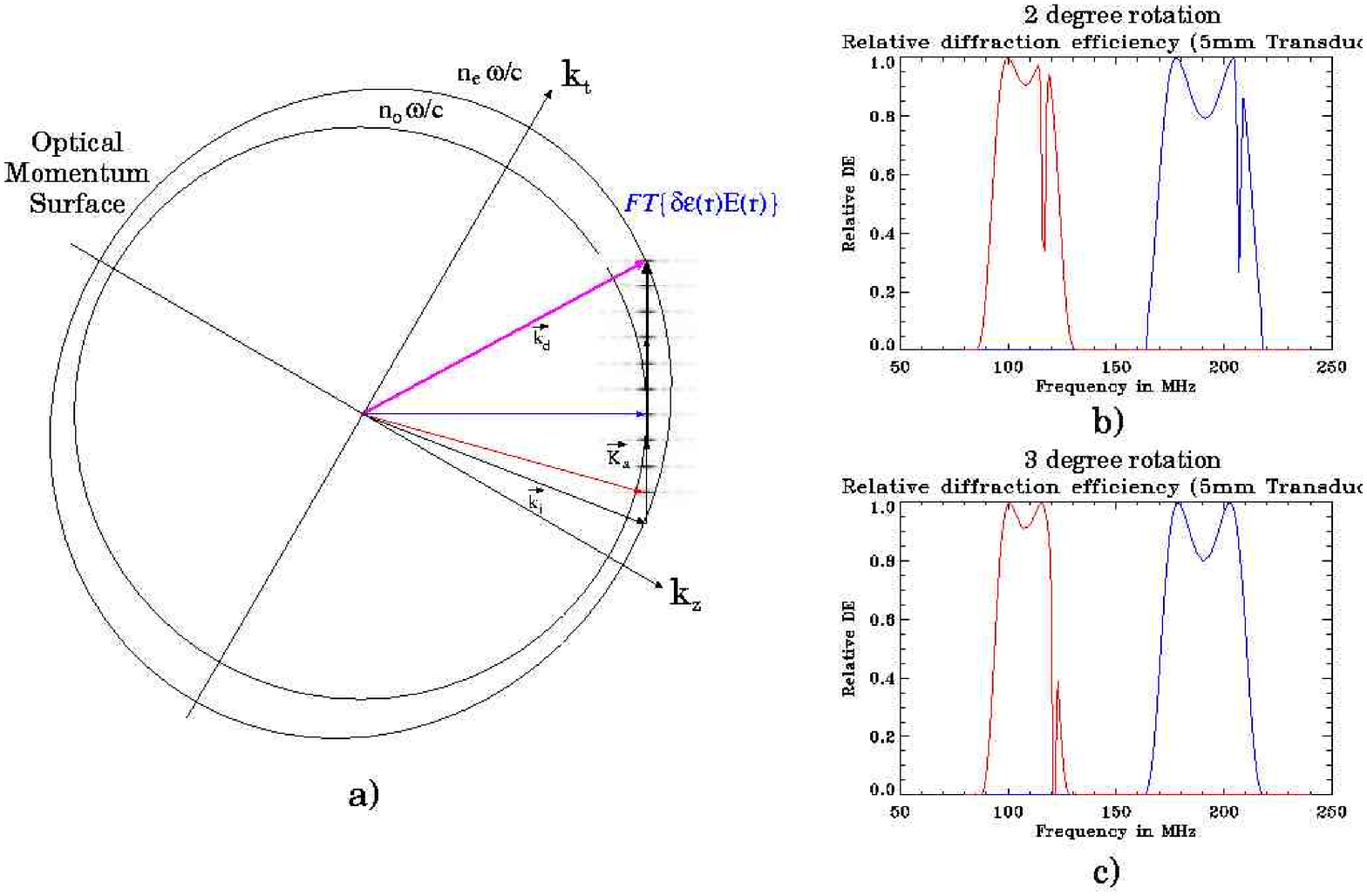}
}
\caption{ a) Optical momentum space showing 
degenerate double diffraction.  
As the applied frequency changes, the acoustic momentum vector 
changes its length proportionally and diffracts light from the 
extraordinary outer to the ordinary inner momentum surface.  
However, part of this diffracted light can rediffract from 
the inner to the outer optical momentum surface (degenerate diffraction) 
unless sufficient acoustic rotation away from the symmetric condition breaks 
the symmetry required for this double diffraction.    
b) With two degrees of acoustic rotation, the degenerate diffraction is 
inside our usable bandwidth.  
c) With three degrees of acoustic rotation, the degenerate 
diffraction dip is just outside our usable bandwidth. }
\label{fig:degenkspc}
\end{figure}  

With acoustic rotation, the midband degeneracy can be eliminated by 
breaking the symmetry of the second order diffraction, and the center 
frequency can be tuned as well~\cite{Yano:75,Maak:99}.  
With insufficient acoustic rotation the in-band degenerate diffraction 
causes a nonlinear dip in the RF bandshape of the tangential Bragg-matched 
diffraction bandshape, as shown in Figure~\ref{fig:degenkspc} a), 
which is especially observable at high diffraction efficiency.  
For an insufficient amount of acoustic rotation, the acoustic momentum 
vector of length $|\vec{K}_{a}|=2\pi f/V_{a}$ can still satisfy 
the momentum-matching condition for light diffraction from the inner 
to outer optical momentum surface, thus rediffracting part of the 
diffracted light (a tangentially Bragg-matched diffraction from the outer 
to inner surface) as shown in Figure~\ref{fig:degenkspc} a).
With further acoustic rotation 
(up to three degrees in the 10 degree optically rotated plane of incidence), 
the in-band degenerate diffraction is completely outside the usable 
bandwidth at both wavelengths as shown in Figure~\ref{fig:degenkspc} c).

\begin{figure}
\centerline{  
 \includegraphics[width=7.0cm, angle=-90]{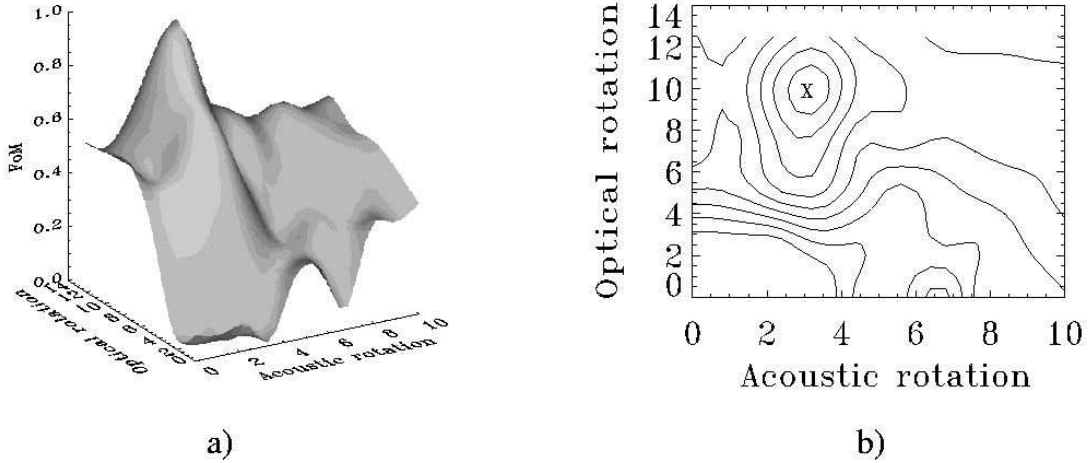}
}
\caption{ The figure-of-merit surface as a function of acoustic 
rotation and optical rotations in degrees. b) The topview of the 
figure-of-merit surface showing optimum point at 10-degree optical 
rotation and three-degree acoustic rotation.  }
\label{fig:opt}
\end{figure}

However, acoustic rotation produces a large walk-off angle that 
increases at $b_{z}$=11 times the rotation angle, which increases the 
required crystal volume.  
Also, the bandshapes for both wavelengths must be 
completely separated and fit within a one-octave bandwidth because 
over an octave the acoustic and transducer nonlinearities may yield 
undesired diffractions. 
This is just barely achievable in this case since the wavelength ratio 
is $\frac{780}{480}=1.62$, thus the difference between center frequencies 
for 780 nm and 480 nm diffraction are inside one-octave bandwidth and 
the fractional bandwidth at both is about 0.2, making each bandwidth for 
780 nm and 480 nm not to exceed an overall one-octave bandwidth.
The high-frequency end is limited by the available transducer 
impedance-matching technology to about 230 MHz, and in addition the 
frequency dependence of the acoustic absorption (18dB/$\mu$s/GHz$^{2}$) 
requires that high frequencies are avoided if high resolution is required.

\begin{figure}
\centerline{
 \includegraphics[width=6.2cm, angle=-90]{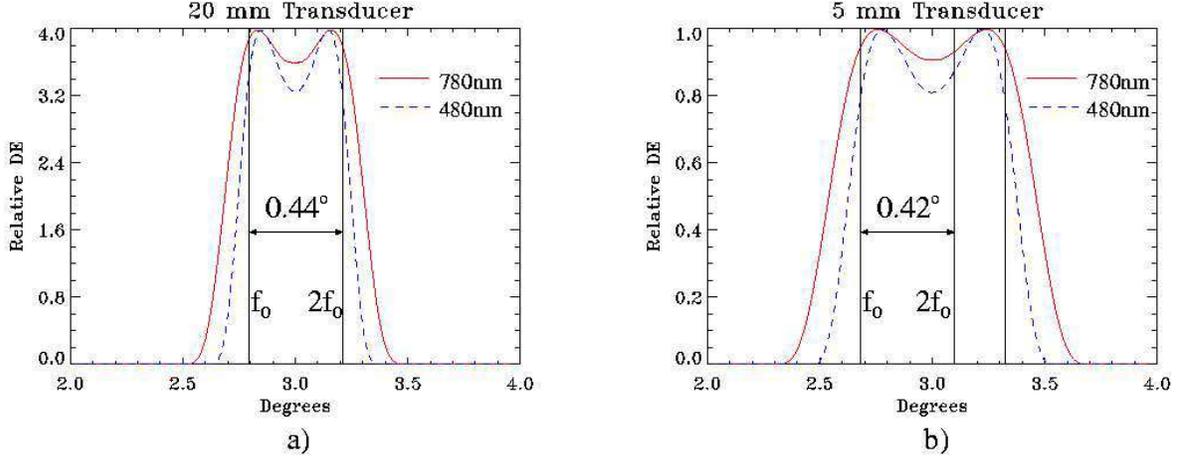}
}
\caption{ a) The bandshape of the designed 
AO device with a 20 mm transducer length shown in the angular domain and 
demonstrating good overlap between the two wavelengths (780/480 nm) and 
centered at three degrees (the acoustic rotation angle).  
780 nm bandshape is plotted with 
solid line and 480 nm bandshape is plotted with dotted line.
Octave bandwidth limits are delimited by the vertical lines.  
b) Since such a large transducer is difficult to 
impedance match, a smaller 5 mm-long transducer was evaluated and is 
shown to have wider bandwidth and angular scan width.  
However, the delineated octave bandwidth limit is about the same as 
the 20 mm transducer, although the efficiency is 4 times lower.}
\label{fig:trleng}
\end{figure}

Considering all these limitations, various optical and acoustic 
rotations were evaluated to find the optimized crystal orientation of 
the AOD in order to find a large bandwidth for both wavelengths 
(thus giving more resolvable spots) while minimizing the bandwidth 
in excess of an octave, $\rm{BW_{ovoctv}}$, as well as the crystal size.
A figure of merit that increases with the bandwidth at both 480 nm and 
780 nm but is penalized for large crystal volume and bandwidth in excess 
of an octave is given in Equation~\ref{eq:fom}:
\begin{equation}
\rm{Figure}\hspace{0.1cm} \rm{of}\hspace{0.1cm} \rm{Merit}=\frac{BW_{\rm{blue}}+BW_{\rm{red}}-BW_{\rm{ovoctv}}}{\rm{Crystal Volume}} \cdot U_{\rm{hflimit}} \cdot F_{\rm{midgen}}, 
\label{eq:fom} 
\end{equation}
where $\rm{U_{hflimit}}$ is a function representing the high frequency 
limit due to the constraints of the transducer impedance-matching 
technology (around 230 MHz), 
and $\rm{F_{midgen}}$ denotes a function penalizing a geometry with the 
degenerate diffraction inside the usable bandwidth.      
This figure of merit is plotted in Figure~\ref{fig:opt} as a function 
of optical and acoustic rotation angles and was used to choose an 
optimized device geometry.  This shows that with a 
10-degree optically rotated plane of incidence and three-degree acoustic 
rotation of the transducer face, the optimized performance is achieved. 
Figures~\ref{fig:trleng} a) and b) show the bandshapes of the designed 
AOD for both wavelengths plotted in the angular domain.  
The bandshapes for both wavelengths are angularly overlapped and centered 
at 3 degrees.  The 20 mm transducer is fully optimized to just fit 
within an octave bandwidth as limited by the red wavelength on the low 
frequency (angle) side and the blue wavelength on the high 
frequency (angle) side.  
The undesired diffraction from the second order acoustic nonlinearity or 
multiple diffraction is still inside the usable bandwidth for 
the 5 mm-long transducer 
(shown as vertical lines in Figure~\ref{fig:trleng} b)), but the 
usable bandwidth is nearly the same as for the 20 mm-long transducer.  
Considering that the 20 mm transducer requires a much larger crystal 
and is much harder to impedance match over a full octave due to the 
large area and capacitance than the 5 mm transducer, we chose to use 
the 5 mm-long transducer.  
There is a decrease in diffraction efficiency in 
\%/Watt (as shown in Equation~\ref{eq:deff}) with the 5 mm 
transducer, but we can still expect more than 80\% diffraction 
efficiency with sufficiently high RF power.  The 480 nm wavelength is 
designed with a 1.0 dB ripple and the 780 nm with a 0.5 dB ripple, so 
that two cascaded devices will have 2dB/1dB ripple, respectively.  
Even though the 480 nm wavelength has a wider bandwidth than the 780 nm 
wavelength in the RF frequency domain, the angular scan range is smaller 
at 480 nm than 780 nm wavelength, as shown in Figure~\ref{fig:trleng}.  
Since our design goals are maximizing both the RF bandwidth and 
the angular overlap range for both wavelengths, the 480 nm wavelength 
was designed with more ripple (giving wider bandwidth) than was used at 
the 780 nm wavelength.  

\begin{figure}
\centerline{
 \includegraphics[width=5.2cm, angle=-90]{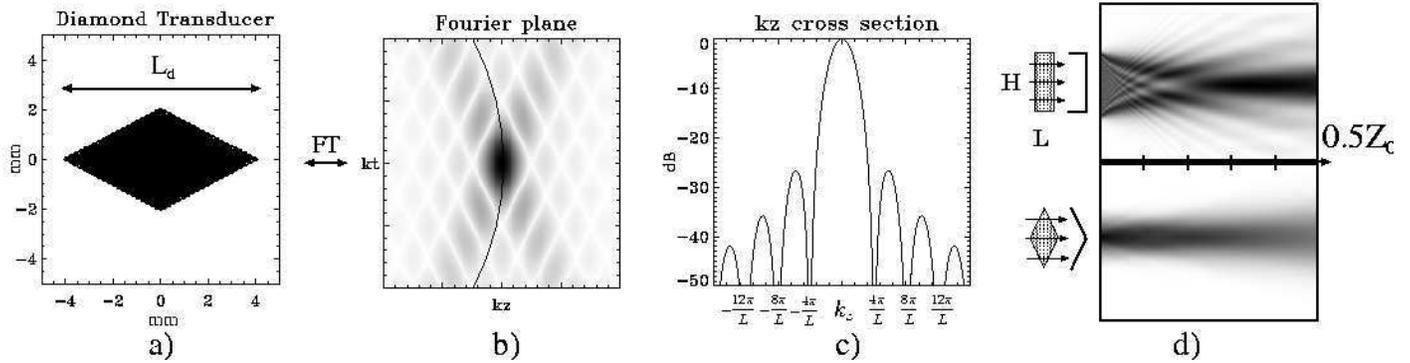}
}
\caption{ a) Diamond-shaped transducer with 8 mm length 
and 4 mm height.  b) Fourier transform of diamond-shaped transducer and 
Bragg-matched loci due to the momentum surface intersections.  
c) k$_{z}$-dimensional cross section of the Fourier plane.  
d) Beam propagation of rectangular and diamond-shaped transducer.  }
\label{fig:diatrans}
\end{figure}

Once the transducer length $L$ has been chosen, the height $H$ can be 
selected in order to keep the full aperture $A$ in the acoustic 
near field and minimize non-uniformities due to acoustic diffraction.  
The enormous out of plane acoustic curvature, $b_{t}$=52, scales the 
near-field distance for a rectangular transducer back to 
$Z_{0}=\frac{H^{2}}{b_{t}\Lambda_{a}}$ and $H$ is usually chosen so that 
$A<Z_{0}$ to maintain well-collimated acoustic column.  
Alternatively, a diamond or truncated diamond transducer 
can be used to apodize the acoustic propagation, thereby producing a more 
uniform acoustic field projection as shown in Figure~\ref{fig:diatrans} d), 
which is a projection of the 3-D propagated acoustic field.  
In the orthogonal Bragg plane, 
the sidelobes of the Bragg selectivity are lowered by the triangular 
apodization of the projection of the diamond, which lowers the Bragg 
mismatched crosstalk.  We consider an 8mm$\times$4mm diamond-shaped 
transducer as shown in Figure~\ref{fig:diatrans} a) as an alternative 
to a 5mm$\times$3mm rectangular transducer with the same area and 
capacitance.
The Fourier distribution of the diamond transducer acoustic wave, 
which is the amplitude weighting painted across the momentum surface 
(as illustrated in Figure~\ref{fig:teo2} a)), 
is shown in Figure~\ref{fig:diatrans} b) along with the curved 
intersection between the optical and acoustic momentum surfaces 
at midband.  
The advantage of using a diamond-shaped transducer over a rectangular 
transducer is the small amplitude of the in-plane first order sidelobe 
(-26 dB compared with -13 dB for the rectangular transducer).  
The in-plane uncertainty distribution of the diamond shaped transducer 
is $\rm{sinc}^{4}\it{(k_{z}\frac{L_{d}}{\rm{4\pi}})}$, 
giving an RF bandshape with a narrower 3 dB bandwidth and lower sidelobe 
compared with the rectangular transducer 
($\rm{sinc}^{2}\it{(k_{z}\frac{L_{s}}{\rm{2\pi}})}$) where $L_{d}=2L_{s}$, 
as shown in Figure~\ref{fig:diatrans} c).  
Figure~\ref{fig:diatrans} d) shows an acoustic beam propagation 
simulation of the rectangular transducer and the diamond transducer 
corresponding to the projection of the three-dimension acoustic field 
as accumulated by the Bragg-matched read out laser.  
The diamond transducer has a more uniform acoustic intensity 
distribution near the transducer compared with the rectangular transducer.  
These nonuniformities will diffract light toward the vertical $k_{t}$ 
dimension, producing vertical sidelobes, as also visible in the 
$\vec{k}$-space Bragg-matched loci shown in Figure~\ref{fig:diatrans} b).  

\begin{figure}
 \centerline{  
 \includegraphics[width=6.2cm, angle=-90]{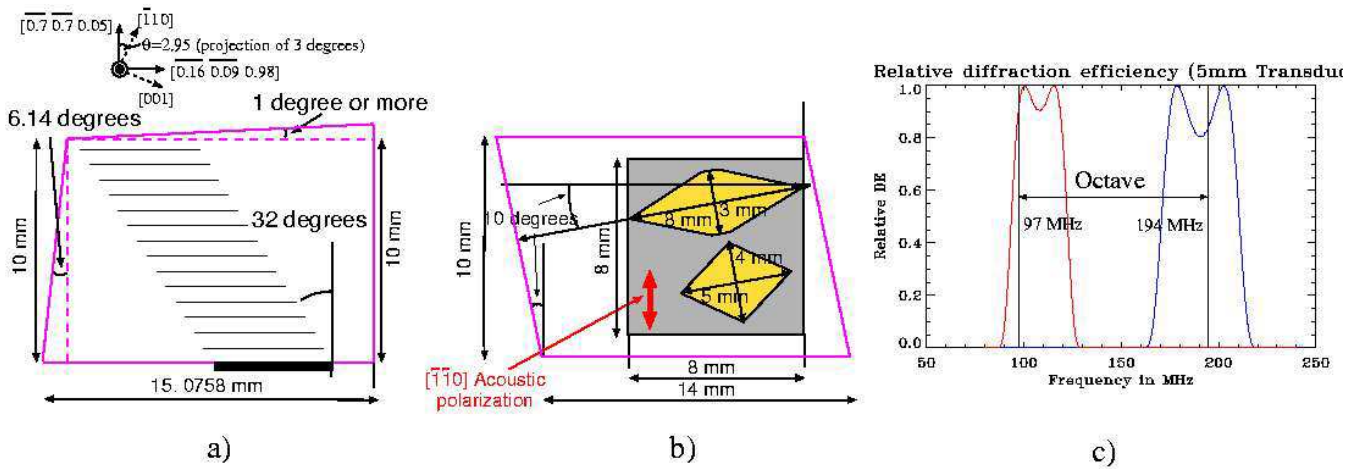}
}
\caption{ a) The top view of the designed AOD 
showing a prism wedge cut for optical input face and acoustic walk-off,  
b) The sideview of the designed AOD showing 10-degree 
optically rotated input face with two diamond-shaped transducer electrodes 
(5 mm and 75 \% truncated 8 mm),  c) The expected bandshape for 
780 nm/480 nm with 5 mm transducer.  }
\label{fig:concl}
\end{figure}

The designed AOD uses a crystal about 1 cm long and 1.5 cm wide.  
The walk-off angle due to three degrees of acoustic rotation is 32 
degrees, as shown in Figure~\ref{fig:concl} a).  
The front face has a 6.14-degree prism cut to satisfy the Bragg-matching 
condition of collinearly incident 780 nm and 480 nm beams. 
The upper face has a one-degree wedge to Bragg-mismatch acoustic 
wave reflections going back to the transducer.  
Figure~\ref{fig:concl} b) shows the 10-degree inclined front surface 
for optical rotation and the inclined diamond transducer electrodes.  
The piezoelectric transducer is oriented with the shear particle 
motion along the $\bar{1}\bar{1}0$ axis and two transducers were 
implemented: an L=5 mm diamond and an L=8 mm truncated diamond transducer. 

%

Unlike unrotated or isotropic AODs in which a single device can be 
rotated to Bragg match either the plus or minus Doppler order, for 
non-symmetric acoustically rotated devices, separate AODs need to be 
fabricated for Doppler upshifting and downshifting interactions 
in order to combine as a Doppler-compensated pair.  
For the AOD with the minus-order diffraction, we need to place the 
transducer at the upper face of the AOD crystal, as shown 
in Figure~\ref{fig:concl} a).
The acoustic wave propagates exactly along the same path as in the 
plus-order device, but the direction of propagation is reversed to 
produce a minus-order Doppler shifted diffraction.
For the 780/480 nm wavelengths, the AOD has 108/191 MHz center frequencies 
and 21/34 MHz bandwidths, as shown in Figure~\ref{fig:concl} c); 
thus the bandshapes (97-119 and 174-208 MHz) are widely separated 
and nearly contained within an octave bandwidth (97-194 MHz).  
The optical aperture is 10 mm in length and 3-4 mm in height, and 
the full aperture access time is 16 $\mu$s, yielding a time-bandwidth 
product of 336 for 780 nm and 560 for 480 nm.
For a circular beam with 4 mm 1/e$^{2}$ diameter, over 100 resolvable spots 
can be achieved with an access time of 4 $\mu$s, which should allow a 
32 qubits, one-dimensional array of well-resolved atoms to be rapidly 
addressed with low crosstalk for this quantum computing application.  
The cascaded AODs also work as two-dimensional scanner, although with a 
Doppler shift given by the sum or difference of the driving frequencies 
along the two axes.  
To address the two-dimensional array of atoms (32 $\times$ 32), the 
Doppler shift can be pre-compensated by acousto-optic modulators (AOM) 
for each wavelength (780/480 nm) driven by the appropriate sum frequency.

\section{Experimental results}\label{sec:result}

In order to characterize the performance of our Doppler-free, 
dual-wavelength scanner, we first made bandshape and diffraction efficiency 
measurements of the AODs with collinear multi-wavelength laser beam 
inputs aligned at the Bragg angle.  We used a 476 nm argon laser and 785 nm 
laser diode for our test light sources since modeling shows that these 
wavelengths are close enough to test the performance of devices 
originally designed for the 780/480 nm wavelengths.

\subsection{Bandshape measurement}\label{sec:bspms}         

\begin{figure}
 \centerline{  
 \includegraphics[width=11.7cm, angle=-0]{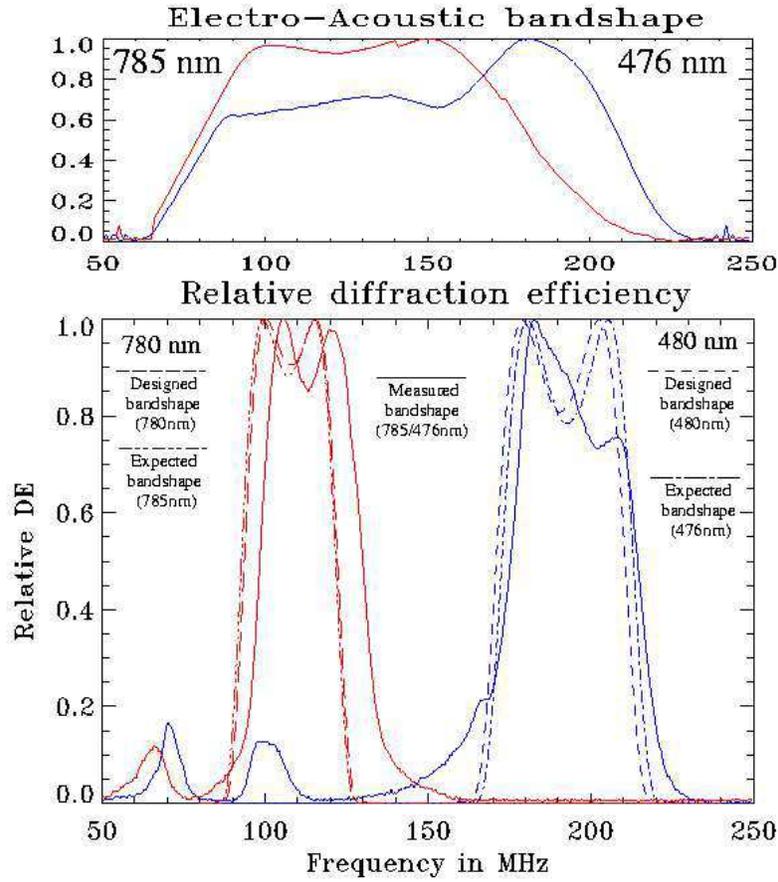}
}
\caption{  Comparison of the designed (780/480 nm) and 
measured (785/476 nm) bandshape of the fabricated AOD.  Bandshapes for 780 nm 
and 480 nm wavelength are well-separated in the RF frequency domain 
and have around 30 MHz of bandwidth.  Theoretical bandshape at 785/476 nm 
is also shown, which is down and up-shifted compared with the 780/480 nm 
design.  The measured electro-acoustic bandshape at 785/476 nm is also 
shown. }
\label{fig:measuredbsp}
\end{figure}  

Both the 476 nm and the 785 nm beams with 4 mm diameter are collinearly 
incident on the AOD with a prism cut at the front face, so the beams 
refract into the crystal at two slightly different angles for appropriate 
Bragg matching designed to give 1 dB ripple at 480 nm and 0.5 dB ripple 
at 780 nm.  The bandshapes are measured with a signal sweeping 
from 50 to 250 MHz.    The applied RF chirp signal drives the AOD, 
which diffracts the collinearly 
incident light (785/476 nm) to the same range of diffracted angles 
(spatially overlapping in the Fourier plane).  
We placed an f=130 mm aplanatic triplet lens after the AOD 
to focus the diffracted beam to a spot that changes its location with the 
applied RF chirp frequency.  
A 1 cm diameter detector head is placed at the Fourier plane to measure 
the diffraction efficiency as the signal generator sweeps in frequency. 
An oscilloscope records the time-varying diffraction efficiency to measure 
the AOD bandshapes.  Figure~\ref{fig:measuredbsp} shows the measured 
bandshapes for both wavelengths, which are well-separated in the frequency 
domain as required for efficient dual frequency operation.  Both bandshapes 
have usable bandwidths of about 30 MHz, which correspond to more than 100 
resolvable spots for a 4 mm diameter incident beam. 
As we rotate the AOD, we can observe the 0.5 dB ripple 
in the 780 nm bandshape.  But for the 480 nm bandshape, the 1 dB ripple 
design is obscured since the bandshape appears to be limited by the 
high-frequency roll-off due to the transducer impedance matching.  
We observed this high frequency behavior by measuring 
the electro-acoustic bandshape, as shown in Figure~\ref{fig:measuredbsp}.  
The electro-acoustic bandshape determines the acoustic power in the crystal 
at a given applied RF frequency by rotating to a perfectly Bragg-matching 
angle for each frequency.  The expected bandshape for 785 nm is slightly 
downshifted from that designed at 780 nm, and the 476 nm is slightly 
upshifted compared to that designed for 480 nm, as shown in 
Figure~\ref{fig:measuredbsp}.  The slight frequency upshift of the measured 
bandshapes at both wavelengths indicates possible cut angle errors.

\subsection{Doppler-free scanning}\label{sec:dfreescan}

\begin{figure}
 \centerline{  
 \includegraphics[width=10.7cm, angle=-90]{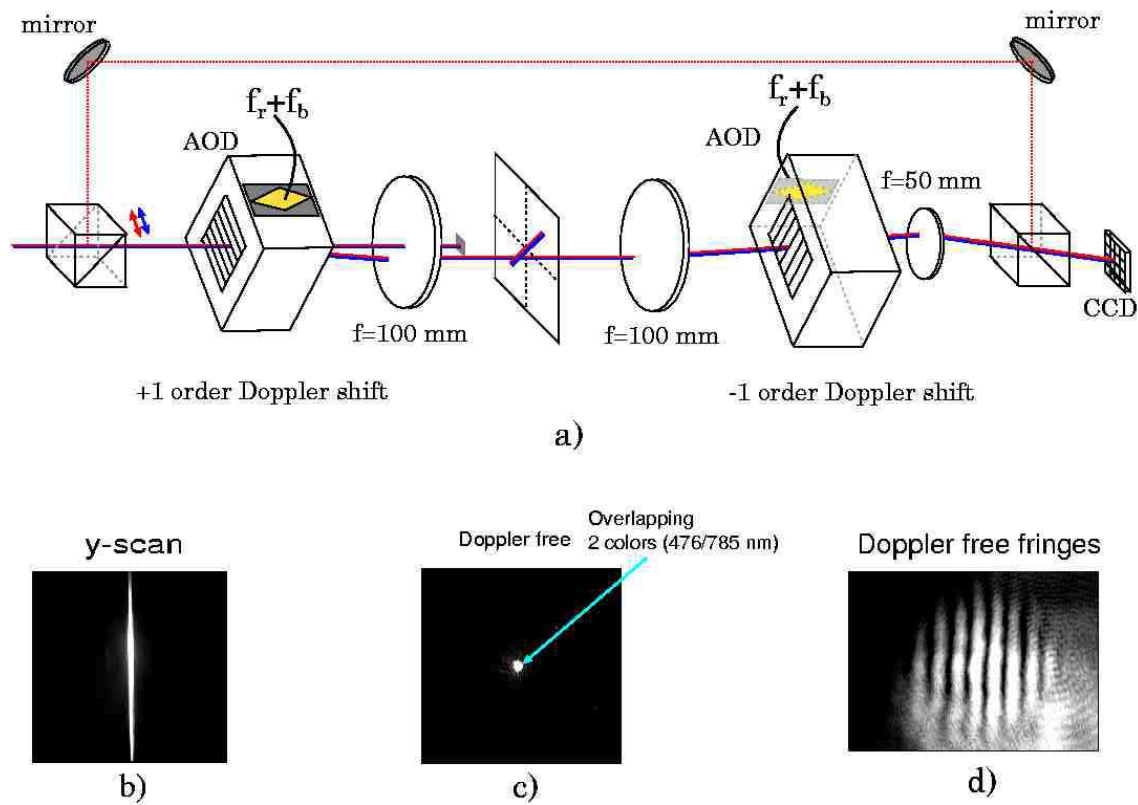}
}
\caption{ Experimental setup and 
dual-wavelength of Doppler-free scanner demonstration.  
a) The first AOD with positive 
Doppler-shifted diffracted light is imaged onto the second 90-degree-rotated 
AOD, which cancels the Doppler-shift with minus order diffraction when driven 
with the same frequency.  
b) The image of the one-dimensional Doppler-free scan 
(y-scan) is shown.   c) Due to the frequency separation of the bandshapes 
for the two colors, there are no undesired diffractions when two different 
single RF frequencies (100/180 MHz) were applied to the AODs.  
d) Doppler-free operation was verified by the stationary 
interference pattern between the laser reference and the doubly diffracted 
beam. }
\label{fig:dfreedemo}
\end{figure}

A linear, Doppler-free scanning experiment was performed with the fabricated 
AODs.  Doppler-free scanning can be accomplished by cascading the diffracted 
light from the first AOD, upshifted by $f$, onto a second AOD driven by the 
same RF frequency, $f$, but oriented to downshift by $f$, yielding net zero 
Doppler-shift, but scanned by both devices.  When the two AODs are in the 
same plane, canceling of the spatial scanning is avoided by having the 
acoustic wave (or images) counter propagate, which doubles 
the diffraction angle.  
But Bragg selectivity in the second AOD, due to 
the different diffraction angles in the first AOD, significantly reduce the 
cascaded bandwidth~\cite{Guilfoyle:PITD,Fukumoto:75}.  
For this reason, we instead cross the two AODs with 
their acoustic wave propagating perpendicular, but with one AOD upshifting 
and the other downshifting, which achieves net Doppler-free scanning along 
the 45$^{\circ}$ bisector oriented vertically, as shown in 
Figure~\ref{fig:dfreedemo}.  
The AOD crystals are placed near the edge of 
their housing for the plus and minus order AODs, so we can place the plus and 
minus AOD crystals right next to each other for Doppler-free cascade 
operation without an intervening telescopic imaging system.  
For our experiments, we did utilize a telescope with two f=100 mm focal 
length achromatic lenses and a DC block.  
We placed our AODs in gimbal mounts with tilt 
(optical rotation) and rotation (Bragg rotation) motion, which are in turn 
mounted in a V-block to rotate the AODs by 45 degrees to give 
combined horizontal (or vertical) Doppler-free scanning.  
Figure~\ref{fig:dfreedemo} b) shows the Doppler-free y-scan captured by a 
Charged Coupled Device (CCD) when both AODs are driven by the same wideband 
chirp and illuminated by both 480 and 780 nm collimated incident light, 
demonstrating cascaded Doppler-free dual-wavelength AO scanner operation.   

Figure~\ref{fig:dfreedemo} c) shows the overlapped two-color Doppler-free 
diffracted spot, achieved using two well-separated RF frequencies 
(100 and 180 MHz) with 10 dBm (10mW) RF power as combined using a power 
adder and amplified with a 20 dB gain-power amplifier giving 50-60\% 
diffraction efficiency for each wavelength (780/480 nm).  
The added RF frequencies were then split into two arms using an RF power 
splitter to drive both of the AODs.  
The separated RF signals diffract each wavelength (780/480 nm) to the same 
location, but due to the well-separated bandshapes (95-115 MHz at 780 nm 
and 170-205 MHz at 480 nm), the 100 MHz 
signal only diffracts the 780 nm beam, and the 180 MHz only diffracts the 
480 nm beam without any undesired crosstalk diffractions.

We set up a Mach-Zehnder interferometer with the undiffracted laser 
directly from the 785 nm laser diode as a reference beam, combined at a 
beamsplitter with the cascaded doubly diffracted beam, to verify that 
the diffracted spot is indeed Doppler-free.  We observed static 
interference fringes, as shown in Figure~\ref{fig:dfreedemo} d), which 
shows that the scanning beam is not Doppler shifted as expected.

\subsection{Small spot addressing}\label{sec:ssaddr}
  
We need to build an optical system to address the $^{87}$Rb atoms 
trapped in closely spaced and tightly focused potential wells for 
Rydberg excitation.  
The size of the potential wells for the $^{87}$Rb atom trap is 
4 $\mu$m and the spacing between trapped atoms is only 8 $\mu$m.   
For successful qubit addressing, the optical system should have 
diffraction-limited performance without any aberrations and small enough 
spots that crosstalk with the neighboring trapped atoms is negligible.  
We used a custom triplet lens that cancels chromatic aberrations and 
the large spherical aberrations induced by the thick vacuum chamber 
windows necessary for atom trapping and cooling (lens designed by 
Thad Walker, University of Wisconsin) .  
The custom triplet lens works as a 3.64$\times$ demagnifier when used 
with an f=400 mm collimation lens.  
To generate spots less than 4 $\mu$m in diameter inside the vacuum chamber, 
we need to appropriately choose the size of the beam incident on the AOD, 
and the focal length of the Fourier lens after the Doppler-free AOD scanning 
system.
The designed heights of the two transducers are either 3 mm or 4 mm for 
each AOD, which limit the maximum size of the spot illuminating the AODs 
since plano-convex cylinder lenses should be avoided when trying to maintain 
diffraction limited performance.  If the Gaussian beam incident on the AOD 
scanner has a diameter $D$=3 mm, we need to use an f=42 mm Fourier lens 
to generate a 13.9 $\mu$m focused spot for 780 nm and an 8.6 $\mu$m 
focused spot for 480 nm, which are ideally then demagnified to 3.8 $\mu$m 
(780 nm) and 2.4 $\mu$m (480 nm) focused spots by the custom triplet in 
combination with the f=400 mm collimation lens.  

\begin{figure}
 \centerline{  
 \includegraphics[width=10.7cm, angle=-90]{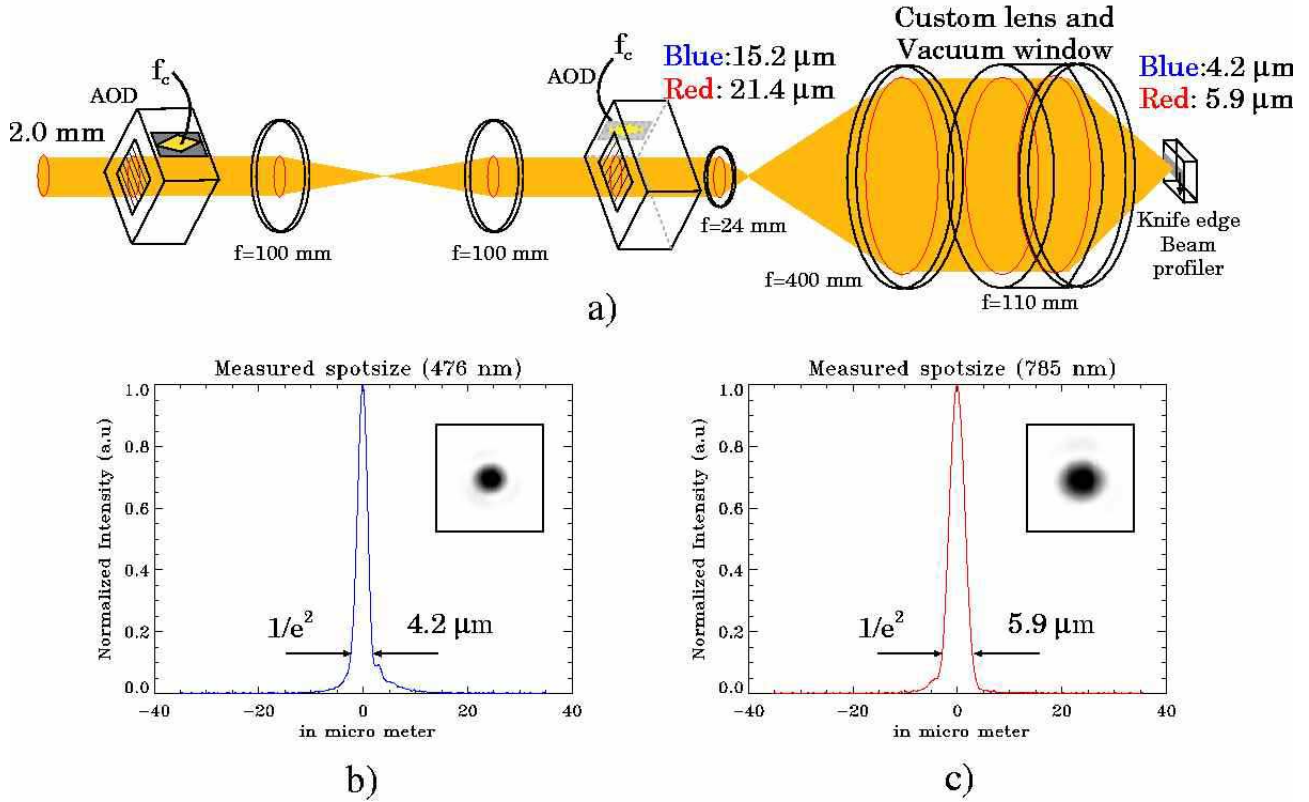}
}
\caption{ a) Experimental setup for measuring 
the achievable spotsize of the cascaded AODs and optical system.   
b) 4.2 $\mu$m 1/e$^{2}$ intensity width of the measured spot at 476 nm 
wavelength.  The inset shows a magnified spot image produced by a 750 mm 
focal length lens placed after the cascaded AODs that is captured by a CCD 
camera and shows low aberration.
c) 5.9 $\mu$m 1/e$^{2}$ intensity width of the measured spot at 785 nm 
wavelength also shows low aberration at 785 nm wavelength. }
\label{fig:spotqual}
\end{figure}  

We experimentally demonstrated the required performance of the Doppler-free 
AOD scanner to address the closely-separated (8 $\mu$m) atoms in the 
quantum information processor with minimal unwanted crosstalk with 
neighboring atoms.  
Figure~\ref{fig:spotqual} a) shows the experimental setup for examining the 
wavefront quality and resulting focal spotsize of our Doppler-free 
AOD scanner.  The Doppler-compensating, 90-degree-crossed AODs were cascaded 
using a telescope imaging system (f=100 mm achromatic doublet lenses).  
The incident Gaussian beam had 2.0 mm 1/$e^{2}$ diameter clipped by a 2 mm 
circular aperture.   The Fourier lens, after successive diffraction by 
both AODs, had a 24 mm focal length.  
A knife edge beam profiler was used to measure 1/$e^{2}$ full width of the 
focused spots (785/476 nm) as 21.4 $\mu$m for 785 nm and 15.2 $\mu$m for 
476 nm. 
The f=400 mm lens collimated the focused spot, and the f=110 mm custom 
triplet lens demagnified the focused spots to 4.2 $\mu$m (476 nm) and 
5.9 $\mu$m (785 nm) 1/$e^{2}$ diameter after the 10 mm thick vacuum window.  
The demagnification ratio is around 3.62 for both wavelengths.
We also placed a 750 mm focal length Fourier lens after the cascaded AODs 
to examine if there is any noticeable distortion in the spot shape after 
cascaded diffraction from the AODs.  The captured spot images 
in the inset of Figures~\ref{fig:spotqual} b) and c) show the 
magnified diffracted spots without significant aberrations.   


Without aberrations and with uniform illumination of the device aperture, 
the number of resolvable spots are given by the time-bandwidth product, 
$T \cdot B$.  As shown in Figure~\ref{fig:measuredbsp}, 
the bandwidth of the fabricated AODs is about 30 MHz.  
However, when the plus and minus AODs were cascaded, the cascaded 
bandwidth was reduced to about 15 MHz.  
The QIP application requires $\mu$s scale spot addressing and switching 
time in order to perform many operations before the decoherence time.  
Assuming a 2 $\mu$s required switching time, which can be achieved using 
a 1.2 mm beam, the number of resolvable spots of the one-dimensional 
Doppler-free scanner is $T \cdot B$=30.    
This indicates that we can address 30 Rydberg atomic qubits, or 10 qubits 
with an over-resolving factor of three for well resolved one-dimensional 
scanning.  
For two-dimensional scanning, we can use the crossed cascaded AODs with 
Doppler pre-compensation achieved using AOMs for both wavelengths.  
We can expect to address well resolved 10 $\times$ 10 two-dimensional 
array of atoms with 2 $\mu$s switching time.

\section{Conclusion}\label{sec:conclu}

We have presented the design, experimental demonstration and characterization 
of a novel AOD, which simultaneously diffracts two 
incident optical wavelengths (780 nm and 480 nm) with the exact same 
diffraction angle by using two proportional frequencies for application in 
two-photon addressing of an array of trapped $^{87}$Rb atoms.
The AOD is designed to use two well-separated Bragg-matched frequency 
subbands within an overall octave RF bandwidth.  
The optimum crystal orientation with both optical and acoustic rotation was 
found by investigating the details of the interaction in momentum space.  
A prism cut was implemented at the optical input surface to allow collinearly 
incident input beams (780 nm/480 nm) to refract into the appropriate 
Bragg-matched incident angles and to give parallel input and diffracted 
output beams at midband, which simplifies the alignment on an optical rail.   
We can address an array of 30 well-resolved atoms rapidly (2 $\mu$s)
with the designed AOD at both 780 and 480 nm, with the diffracted spots 
precisely overlapped in the Fourier plane.
The wavefront quality of the cascaded AOD 
scanner was examined by observing the focused spots after optical imaging  
with an optimized low aberration system through a thick vacuum window.  
For addressing two-dimensional arrays with crossed AODs, we expect to be able 
to address 10 $\times$ 10 arrays of atoms for QIP application by using 
Doppler pre-compensation with an acousto-optic frequency shifter.  
In a subsequent paper, we will show how to simultaneously address 
multiple spots without crosstalk using a new cascaded AOD design. 

For the optimum performance of the cascaded AODs, as well as for addressing 
the atoms with the desired polarizations, we need to be able to manipulate the 
state of polarization for both wavelengths simultaneously incident on the 
AOD to match the crystal eigenmodes 
and to transform the diffracted output polarization (which varies slightly 
with frequency) into the required polarization for the next stage. 
In a companion paper, we will describe the design of bichromatic waveplates 
that transform wavelength dependent eigenmode polarizations for cascaded 
AODs and another bichromatic waveplate to generate the required 
polarizations for Rydberg excitation after the cascaded AODs.

\vspace{12pt}
\section*{Acknowledgements}

The authors gratefully acknowledge the support of the National Science 
Foundation (NSF) ITR program and Warren Seale from NEOS technology for 
helpful discussions about fabricating these AODs. 
This work was supported by the NSF (grant \# PHY-0205236) and ARO-DTO 
(contract \# DAAD19-02-1-0083).

\bibliographystyle{osajnl}

\end{document}